\documentstyle[epsf,eqsecnum,floats,preprint,aps]{revtex}

\tighten

\begin{document}
 
\title{Quantum Effects on Higgs Winding Configurations
\footnote[1]{To be published in the proceedings of A CRM-Fields-CAP
Summer Workshop in Theoretical Physics: Solitons, Properties, Dynamics,
Interactions and Applications, Kingston, Ontario, Canada, 20-26 Jul 1997.}}
 
\author{Arthur Lue\footnote[2]{lue@cuphyb.phys.columbia.edu}}

\smallskip

\address{Department of Physics\\ 
Columbia University \\
New York, NY 10027\\}

\maketitle

\vfill

\begin{abstract}

We examine the quantum corrections to the static energy for Higgs
winding configurations.  We evaluate the
effective action for such configurations in Weinberg-Salam theory
without $U(1)$-gauge fields or fermions.  For a configuration whose
size is much smaller than the inverse W-mass, quantum contributions
to the energy are comparable to the classical energy.  Moreover, it
is insufficient to consider only one-loop corrections, even as
$\hbar\rightarrow 0$.  Indeed, all loop orders contribute equally
to the static energy.  Nevertheless, quantum fluctuations do not
stabilize winding configurations.

\end{abstract}

\setcounter{page}{0}
\thispagestyle{empty}
 
\vfill
 
\noindent CU-TP-865 \hfill

\noindent hep-ph/yymmdd \hfill Typeset in REV\TeX

\eject
 
\vfill
 
\eject
 
\baselineskip 24pt plus 2pt minus 2pt

\section{Introduction}

The Higgs sector in the standard model is a linear sigma model.  Such
a theory exhibits configurations of nontrivial winding, though they
are not stable.  Winding configurations in the standard model shrink
to some small size and then unwind via a Higgs zero when allowed to
evolve by the Euler-Lagrange equations.  These winding configurations
can be stabilized if one introduces four-derivative Higgs
self-interaction terms which are not present in the standard model
\cite{giptze1,giptze2,ambrub}.
The motivation typically cited for introducing
such terms is that one may treat the Higgs sector of the Lagrangian as
an effective field theory of some more fundamental theory which only
manifests itself explicitly at some high energy scale.  The stabilized
configurations have phenomenological consequences in electroweak
processes and provide an arena for testing nonperturbative aspects of
field theory and the standard model.

Because the procedure just described for stabilization is
inconsistent, we will take a different approach; we wish to see
whether just the quantum fluctuations of a {\em renormalizable}
$SU(2)$-Higgs theory can stabilize winding configurations.  We will
take the Higgs sector to be that found in the standard model.
In this paper, we identify the quantum effects on the energy of static
winding configurations by evaluating the effective action.
If quantum
effects stabilize solitons, that effect should be reflected by some
extremum in the effective action.  If we take the weak gauge-coupling
limit, $g^2 \rightarrow 0$, an analytic expression
is available for the effective action.  The weak coupling limit is
equivalent to the semiclassical limit when fields are scaled properly.
When Planck's constant is small, we need only focus on small field
configurations.  It is only for such configurations that quantum
corrections are important, and thus have the possibility of stabilizing
configurations which are unstable classically.

\section{Asymptotic Behavior of the Effective Action}

Consider the Weinberg-Salam theory of electroweak interactions,
neglecting the $U(1)$-gauge fields and fermions.  Our field variables
form the set $\{A_\mu(x),\phi(x)\}$ where the gauge field $A_\mu(x) =
\sigma^aA_{\mu a}(x)/2$ is in the adjoint representation of $SU(2)$
($\{\sigma^a\}$ are the Pauli matrices), and the Higgs field $\phi(x)$
is in the fundamental representation of $SU(2)$.  We choose the
$R_\xi$-gauge to properly quantize this theory.
In the following treatment, the parameter $m$ is the mass of
gauge field (the W-particle) and $m_H$ is the
physical Higgs mass.  The Feynman rules derived from the specified
action are familiar.  In this analysis, we restrict ourselves to the
semiclassical limit, which is equivalent to taking $g^2 \rightarrow 0$
while holding $m,m_H$ fixed.

We wish to determine the effects of quantum fluctuations on Higgs
winding configurations.  We will evaluate the effective action,
$\Gamma[A^a_\mu,\phi]$, where  $A^a_\mu(x) = 0$ and
$\phi(x) = \left[U({\bf x}) - 1\right]\phi_0$.  Here, $\phi_0$ is
some constant field such that
$\phi_0^\dagger\phi_0 =m^2/g^2$ and $U({\bf x}) \in SU(2)$ is
a static configuration such that
$U({\bf x}) \rightarrow 1$ as $|{\bf x}| \rightarrow \infty$ with
characteristic size, $a$.  We require the field $U({\bf x})$ to be a
configuration of unit winding number.

The static energy for the state whose expectation value of the operator
associated with the Higgs field is $\phi({\bf x})$ will be the quantity $E$ in
the expression $\Gamma[\phi] = -\int dt\ E$.
The effective action $\Gamma[\phi]$ is the generating functional for the
one-particle irreducible
green's functions with $n$ external $\phi$'s,  $\Gamma^{(n)}$.
Normally, the effective action would not be solvable exactly.
However, because we are investigating the semiclassical limit, we are only
interested in configurations whose size, $a$, is small.

We find that under such a circumstance, we may use the Callan-Symanzik
equation for our theory to evaluate the leading-order size dependence
of the one-particle irreducible green's functions
and thus evaluate the quantum corrections to the static energy.
We implement the condition of small, static background configurations
by requiring the field $\hat\phi(p)$,the Fourier transform of the
field $\phi(x)$,to have support only for $p_0 =
0$ and $|{\bf p}| \gg m^{-1}, {m_H}^{-1}$ which implies $0 <
m^{-2},{m_H}^{-2} \ll -p^2$.  Under this circumstance, the asymptotic
dependence of $\Gamma^{(n)}$ will be determined by the Callan-Symanzik
equation.  The one-loop beta functions may be easily
obtained from the literature \cite{GW}.  The one-loop anomalous
dimension is also easy to evaluate.

So long as $m_H/m$ is not too large,
we find that the leading-order size dependence of the effective
action comes from the two-point one-particle irreducible green's
function.  All other terms are supressed by powers of $a$ and other
factors.  The leading-order contribution to the effective action from
quantum fluctuations yields
\begin{equation}
	\Gamma[\phi] = \int\frac{d^4p}{(2\pi)^4}\phi^\dagger(p)\phi(p)
		p^2\left[1+\frac{1}{2}b_0g^2\ln\left(\frac{-p^2}{m^2}\right)
		\right]^\frac{c_0}{b_0}.
\label{2-point}
\end{equation}
Here
$b_0 = 43/48\pi^2$ and $c_0 = 3[1+(\xi-1)/4]/16\pi^2$, where $\xi$ is the
gauge parameter ($\xi > 0$).  The scale dependence
of the above expression is
\begin{equation}
	\Gamma[\phi] \sim -\int dt\ \frac{m^2a}{g^2}
		\left[1+b_0g^2\ln\left(\frac{1}{ma}\right
		)\right]^\frac{c_0}{b_0}.
\label{energy}
\end{equation}
One can recover the classical result from (\ref{energy}) by setting
the $g^2$ inside the brackets to zero.  Note that when
$b_0g^2\ln(\frac{1}{ma}) \sim 1$, the quantum corrections to the
energy are as significant as the classical contribution.
Nevertheless, the static energy that corresponds to this effective
action is a monotonically increasing function of the size, $a$, such
that $E(a=0) = 0$.  This would imply that Higgs winding configurations would
shrink to zero size and unwind via a Higgs zero, just as in the
classical scenario.

\section{Discussion}

Let us take a closer look at our expression for the leading contribution to the
effective action (\ref{2-point}).  Expanding in powers of $g^2$ we get
$$
		\Gamma[\phi] = \int\frac{d^4p}{(2\pi)^4}\phi^\dagger(p)\phi(p)
		p^2 + \frac{c_0}{2}g^2\int\frac{d^4p}{(2\pi)^4}
		\phi^\dagger(p)\phi(p)p^2\ln\left(\frac{-p^2}{m^2}\right)
		+ \cdots
$$
The first term is the contribution from the classical action.  The next term is
the leading order contribution from one-loop one-particle irreducible graphs.
The scale dependence of the static energy goes like
\begin{equation}
	E = \frac{m^2a}{g^2}\left[{\cal A} + {\cal B}g^2\ln\frac{1}{ma}
		+ {\cal C}g^4\left(\ln\frac{1}{ma}\right)^2 + \cdots\right]
\label{expansion}
\end{equation}
where $\cal A, \cal B, \cal C$ are numbers.  Again the first term is
the classical energy, the second is the one-loop energy, and the rest
of the terms in the expansion (\ref{expansion}) correspond to
higher-loop energies order by order.  We can see by comparing
(\ref{energy}) with (\ref{expansion}) that loop contributions to the
effective action beyond one loop can only be neglected when
$b_0g^2\ln(1/ma) \ll 1$.  However, that is precisely the condition
where the one-loop contribution can be neglected relative to the
classical action.  Thus, drawing conclusions concerning solitons based
on one-loop results may be difficult.  When dealing with small
configurations, one still needs to include higher-loop contributions,
even in the semiclassical limit.

There are limitations to (\ref{2-point}) which we will not discuss here.
Complications occur from $m_H/m$ dependence and the running of the Higgs
self-coupling.  For a more complete discussion, please refer to
\cite{original}.

\acknowledgments

The author wishes to express gratitude for the help of E. Farhi in this work.
The author also wishes to acknowledge helpful conversations with J. Goldstone,
 K. Rajagopal, K. Huang, K. Johnson, L. Randall, M. Trodden, and T. Schaefer.
Moreover, I would also express gratitute to the organizers of this workshop.
This work 
was supported by funds provided by the
U.S.~Department of Energy.


\begin{thebibliography}{}
 
\bibitem{giptze1}
J. M. Gipson and H. C. Tze, Nucl. Phys. {\bf B183}, 524 (1981).
\medskip

\bibitem{giptze2}
J. M. Gipson, Nucl. Phys. {\bf B231}, 365 (1984). 
\medskip
 
\bibitem{ambrub}
J. Ambjorn and V. A. Rubakov, Nucl. Phys. {\bf B256}, 434 (1985).
\medskip
 
\bibitem{GW}
D. J. Gross and F. Wilczek, Phys. Rev. D {\bf 8}, 3633 (1973).
\medskip

\bibitem{original}
A. Lue, Phys. Rev. D {\bf 55}, 6725 (1997).

\end{thebibliography}
\end{document}